\documentclass[a4paper,11pt]{article}
\usepackage{pos}
\usepackage{amsmath}
\usepackage{placeins}

\title{Exploring new directions in enhancing the ACTS parameter optimization suite}
\ShortTitle{Exploring new directions in enhancing the ACTS parameter optimization suite}
\renewcommand{\printHeadAuthors}{LaVoie et al.}

\author*[a]{Chance LaVoie}
\author[b,c]{Qi Bin Lei}
\author[c]{Rocky Bala Garg}
\author[c]{Lauren Tompkins}

\affiliation[a]{Department of Mechanical Engineering, Carnegie Mellon University,\\
  Pittsburgh, PA 15213, USA}

\affiliation[b]{Department of Physics, University of California, Santa Cruz,\\
  and Santa Cruz Institute for Particle Physics,\\
  1156 High St., Santa Cruz, CA 95064, USA}

\affiliation[c]{Department of Physics, Stanford University,\\
  Stanford, CA 94305, USA}

\emailAdd{chancel@cmu.edu}
\emailAdd{qlei6@ucsc.edu}
\emailAdd{rbgarg@stanford.edu}
\emailAdd{laurenat@stanford.edu}

\abstract{Track seeding strongly affects both the quality and computational cost of charged-particle reconstruction, yet its many configuration parameters are commonly tuned through expert intuition and repeated trial and error. ACTS reduces this burden with an Optuna Tree-structured Parzen Estimator auto-tuner, but expensive evaluations, a restricted search space, and a scalarized objective can limit evaluation efficiency, exclude promising configurations, and obscure performance trade-offs.

We investigate whether Bayesian optimization can address these limitations using ACTS with the Open Data Detector (ODD). Under identical search ranges and a common 100-trial budget, we compare Expected Improvement and Upper Confidence Bound with TPE and random search on the existing eight-parameter problem, extend the best-performing Bayesian method to fifteen parameters, and apply Expected Hypervolume Improvement to optimize efficiency, fake rate, duplicate rate, and runtime without fixed scalar weights. Candidate configurations are evaluated through the full ACTS reconstruction chain and validated on disjoint held-out events. The Bayesian acquisition methods identify strong configurations earlier than TPE, and their advantage persists in held-out validation. Expanding the search further improves performance, while multi-objective optimization reveals competitive non-dominated solutions spanning distinct trade-offs. These results indicate that Bayesian optimization can strengthen ACTS auto-tuning through efficient evaluation, broader parameter searches, and post-hoc expert selection among non-dominated alternatives.}

\FullConference{23rd International Workshop on Advanced Computing and Analysis Techniques in Physics Research (ACAT2025)\\
8--12 September 2025\\
Hamburg, Germany\\}

\begin{document}
\maketitle

\section{Introduction}

Track seeding strongly affects both charged-particle reconstruction quality and downstream computing cost. Seeding algorithm parameters are commonly tuned through expert intuition and repeated trial and error, a time-consuming process that depends on detector geometry and operating conditions~\cite{chatain}.

A Common Tracking Software (ACTS)~\cite{acts}, an open-source reconstruction framework, reduces this burden with an Optuna~\cite{optuna} Tree-structured Parzen Estimator (TPE)~\cite{tpe} auto-tuner that maximizes a scalarized score consisting of efficiency, fake rate, duplicate rate and processing time, over eight seeding parameters. Yet three challenges remain: full-chain trials make inefficient evaluation costly; optimal solutions may be difficult to find in the eight-parameter space yet better configurations may exist when more parameters are considered; and fixed scalarization weights hide viable performance trade-offs behind a single number.

We test Bayesian optimization against these challenges. Expected Improvement (EI)~\cite{ego} and Upper Confidence Bound (UCB)~\cite{gpucb} are compared with TPE and random search on the same eight-parameter problem, then EI is extended to fifteen parameters. Expected Hypervolume Improvement (EHVI)~\cite{ehvi} optimizes the metrics separately and returns a Pareto set for post-hoc selection.
\section{Experimental setup}

\subsection{ACTS, ODD, and evaluation}

We use ACTS with the Open Data Detector (ODD), a realistic public HL-LHC testbed~\cite{odd}. We simulate $t\bar{t}$ events at $\sqrt{s}=14$~TeV with 200 pile-up interactions. Each event is processed by the full ACTS chain with a parameter candidate vector $\mathbf{x}$. Using a fixed ten-event training sample, we average runtime $t(\mathbf{x})$ and the CKF-stage efficiency $\epsilon(\mathbf{x})$, fake rate $f(\mathbf{x})$, and duplicate rate $d(\mathbf{x})$. The CKF-stage metrics are evaluated before downstream ambiguity resolution; however, CKF seed de-duplication is enabled. 

\subsection{Optimization methods}

All methods use the same sequential propose--evaluate--update loop. At iteration $n$, the optimizer selects a seeding vector $\mathbf{x}_n$; ACTS returns $(\epsilon,f,d,t)$ or its scalarized score, and the observation is appended to the trial history $\mathcal{D}_n$. TPE refits densities to better and worse observations~\cite{tpe}, whereas Bayesian optimization refits a probabilistic surrogate and scores candidates with an acquisition function, $a(\mathbf{x})$. It then selects $\mathbf{x}_{n+1}=\arg\max_{\mathbf{x}}a(\mathbf{x})$. Thus, both the surrogate and acquisition optimum are updated after every ACTS evaluation. Xopt v2.4 implements this Bayesian loop~\cite{xopt}, which repeats until the trial budget is exhausted.

\textbf{Single-objective optimization:} Single-objective methods require the metrics to be scalarized. We use the ACTS auto-tuning score~\cite{gargexplore,gargsuite}, so all methods maximize the same objective:
\begin{align}
S(\mathbf{x})&=\epsilon(\mathbf{x})-\left[f(\mathbf{x})+d(\mathbf{x})/5+t(\mathbf{x})/5\right]. \label{eq:score}
\end{align}
The Bayesian surrogate provides a predicted score $\mu(\mathbf{x})$ and uncertainty $\sigma(\mathbf{x})$. EI estimates the gain over the best observed score $S^\star$; UCB rewards predicted score and uncertainty:
\vspace{-0.5\baselineskip}
\begin{align}
a_{\mathrm{EI}}(\mathbf{x})&=(\mu-S^\star)\Phi(z)+\sigma\phi(z),\quad z=(\mu-S^\star)/\sigma, \label{eq:ei}\\[-2pt]
a_{\mathrm{UCB}}(\mathbf{x})&=\mu(\mathbf{x})+\beta\sigma(\mathbf{x}). \label{eq:ucb}
\end{align}
Here $\Phi$ and $\phi$ are the standard-normal cumulative distribution and density; $\beta$ controls exploration, and we use $\beta=2$ in this work.

\textbf{Multi-objective optimization:} Multi-objective Bayesian optimization treats efficiency, fake rate, duplicate rate, and runtime as separate objectives rather than combining them into the scalarized score in Eq.~\eqref{eq:score}. EHVI favors candidate configurations predicted to improve the current set of best trade-off solutions. This allows the operating point to be selected after comparing the resulting trade-offs of the optimization targets.

\subsection{Experiments}

Each experiment targets one Section~1 limitation using the Section~2.1 evaluation chain. Within each comparison, all optimizers use the same initialization, bounds, and event samples.
\begin{sloppypar}
\textbf{Evaluation efficiency:} The first experiment compares how efficiently different optimizers use the fixed trial budget of 100 to tune ACTS seeding parameters. The eight parameters are {\footnotesize\texttt{maxSeedsPerSpM, deltaRMin, deltaRMax, impactMax, sigmaScattering, radLengthPerSeed, cotThetaMax, maxPtScattering}}. We run TPE, EI, UCB, and a uniform-random baseline for 100 single-objective trials. We measure convergence by running-best score and each method's best configuration is tested on validation events.
\end{sloppypar}

\textbf{Search-space scaling:} To address the limitations of the size of the parameter set, the second experiment extends the problem from eight to fifteen seeding parameters using EI, adding {\footnotesize\texttt{rMin, rMax, zMin, zMax, collisionZMin, collisionZMax, minPt}}. Across 200 single-objective trials, we test whether the expanded set yields higher scores.

\textbf{Optimization without scalarization:} The third experiment explores tuning without scalarization. Over 250 trials, EHVI treats the four metrics as separate objectives while tuning the original eight parameters. Representative Pareto-front configurations are validated for post-hoc expert selection.

\section{Results}

\begin{figure}[!ht]
\centering
\begin{minipage}[t]{0.49\textwidth}
  \vspace{0pt}\centering
  \includegraphics[height=0.19\textheight]{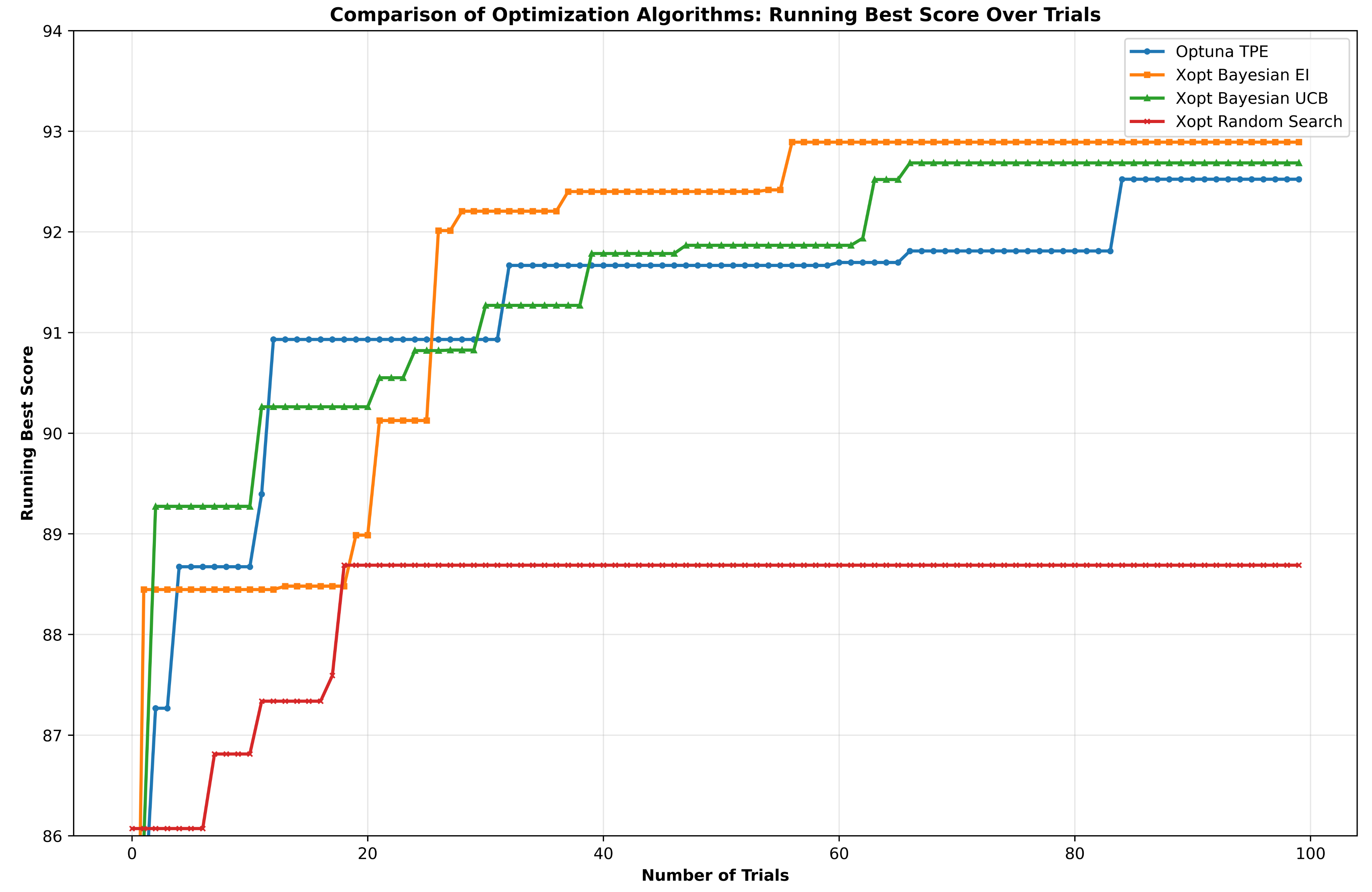}
  \caption{Running-best scalarized score on training events.}
  \label{fig:single-objective}
\end{minipage}\hfill
\begin{minipage}[t]{0.49\textwidth}
  \vspace{0pt}\centering
  \includegraphics[height=0.19\textheight,trim=0 430 0 0,clip]{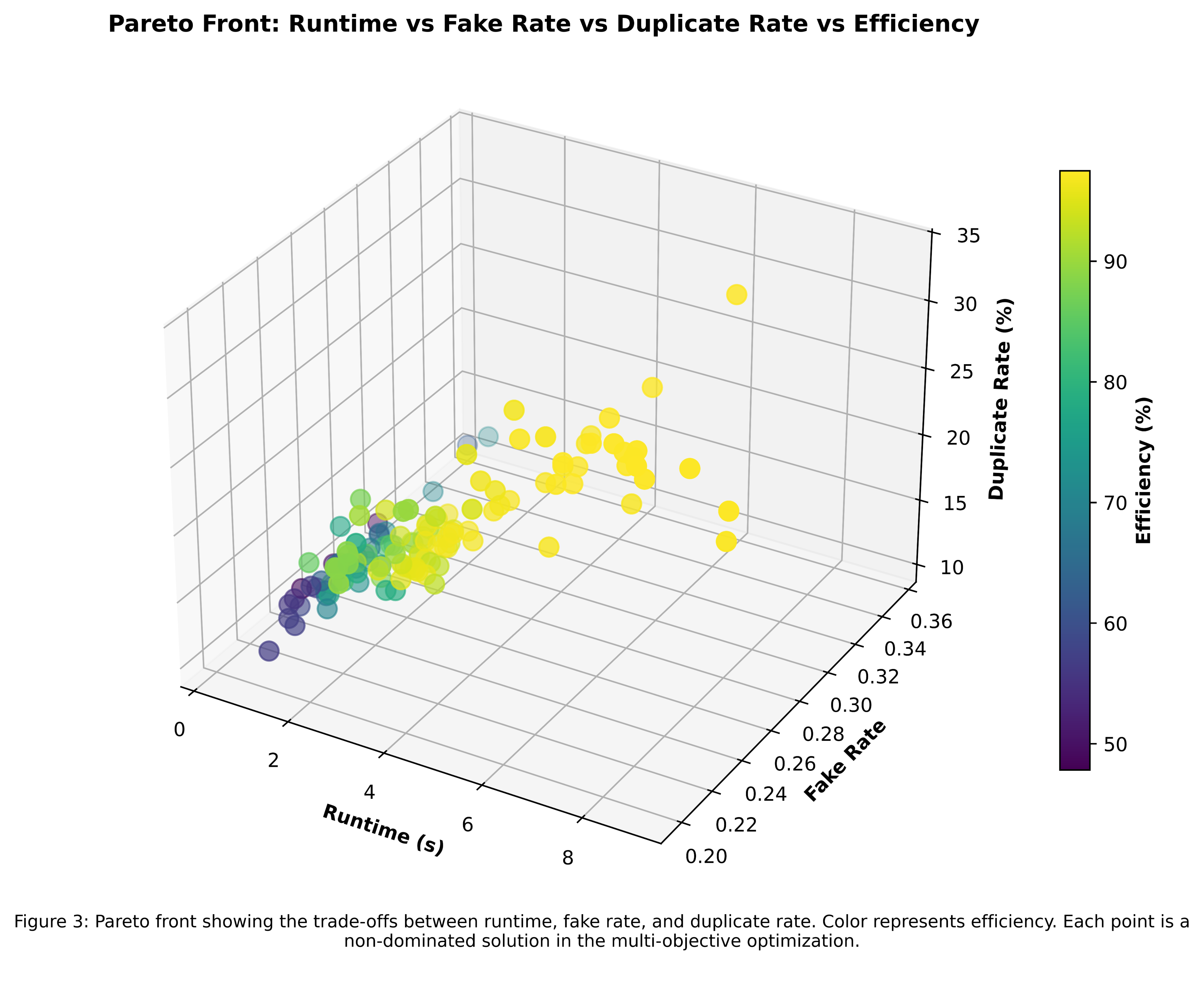}
  \caption{Non-dominated solutions on the training events in the runtime, fake rate, duplicate rate space; color represents efficiency.}
  \label{fig:pareto}
\end{minipage}

\medskip

\begin{minipage}[t]{0.49\textwidth}
  \vspace{0pt}\centering
  \scriptsize
  \resizebox{\linewidth}{!}{%
  \begin{tabular}{lrrrrrr}
  \hline
  Method & $N_p$ & $S$ & $\epsilon$(\%) & $f$(\%) & $d$(\%) & $t$(s) \\
  \hline
  TPE & 8 & 91.99 & 95.80 & .250 & 15.25 & 2.551 \\
  EI & 8 & 92.29 & 96.19 & .252 & 15.51 & 2.735 \\
  UCB & 8 & 92.13 & 96.03 & .252 & 15.52 & 2.730 \\
  Random & 8 & 89.67 & 96.57 & .321 & 27.29 & 5.609 \\
  EI & 15 & 92.73 & 96.62 & .249 & 16.83 & 1.403 \\
  \hline
  \end{tabular}}
  \captionof{table}{Performance on validation events. Number of optimized parameters ($N_p$); single-objective score ($S$); $\epsilon$, $f$, $d$, and $t$ are defined in the text.}
  \label{tab:validation}
\end{minipage}\hfill
\begin{minipage}[t]{0.49\textwidth}
  \vspace{0pt}\centering
  \scriptsize
  \begin{tabular}{lrrrr}
  \hline
  Solution & A & B & C & D \\
  \hline
  $\epsilon$ (\%) & 96.6 & 96.7 & 96.3 & 96.7 \\
  $f$ (\%) & .26 & .27 & .26 & .25 \\
  $d$ (\%) & 14.7 & 15.8 & 14.6 & 16.8 \\
  $t$ (s) & 2.8 & 3.1 & 2.3 & 3.6 \\
  \hline
  \end{tabular}
  \captionof{table}{ $\epsilon$, $f$, $d$, and $t$  on validation events for four configurations selected from the training Pareto front.}
  \label{tab:pareto}
\end{minipage}
\end{figure}

\textbf{Evaluation efficiency on the eight-parameter problem:} Within the common 100-trial budget, EI is first to reach its best observed training score followed by UCB, and then TPE (Figure~\ref{fig:single-objective}). Validation scores reflect the same ordering: 92.29 for EI, 92.13 for UCB, and 91.99 for TPE (Table~\ref{tab:validation}). Earlier convergence to higher-scoring configurations shows that the Bayesian acquisition methods use the limited evaluation budget more efficiently than TPE. All three optimizers outperform the random baseline.

\textbf{Scaling to fifteen parameters:} The fifteen-parameter EI optimization achieves a validation single-objective score of 92.73, an improvement of 0.44 over the best eight-parameter result (Table~\ref{tab:validation}). The selected configuration increases efficiency by 0.43 percentage points and reduces runtime by 48.7\%, with the duplicate rate rising from 15.51\% to 16.83\%. The higher score indicates that additional optimization power is obtained by expanding the range of considered parameters, which is challenging in the non-Bayesian methods due to their relative inefficiency.

\textbf{Optimization without scalarization:} Without fixing metric weights in advance, EHVI identifies a Pareto front of attainable solutions spanning distinct trade-offs (Figure~\ref{fig:pareto}). Evaluation on validation events of representative solutions shows that they remain competitive with single-objective results (Tables~\ref{tab:validation} and~\ref{tab:pareto}). Experts can therefore choose a preferred operating point after observing the alternatives.

\FloatBarrier
\vspace{-1.5em}
\section{Discussion and Conclusion}

Together, the studies address all three challenges: Bayesian acquisition uses a fixed budget more efficiently, fifteen parameters lead to a higher score, and EHVI preserves trade-offs hidden by scalarization. Scalar optimization suits known priorities; EHVI supports selection after exploration.


In conclusion, these results indicate that Bayesian optimization can strengthen ACTS seeding auto-tuning through efficient evaluation, larger searches that reveal better configurations, and Pareto fronts that support expert selection among physics and computing priorities.

\acknowledgments
Code development was co-piloted with OpenAI Codex. OpenAI ChatGPT was used to assist in editing these proceedings. This work was supported by the CMU SIEF Scholarship and NSF Cooperative Agreements OAC-1836650 and PHY-2323298. Computing resources were provided by CERN.

\end{document}